\documentclass[aip,reprint,apl,numerical]{revtex4-1}
\usepackage{graphicx}%
\usepackage{dcolumn}%
\usepackage{bm}%
\usepackage{tabularx}

\usepackage{amsmath}
\usepackage{graphicx}
\usepackage{epsfig}
\usepackage{tabularx}
\usepackage{color}

\begin{document}
\hyphenation{mo-des}
\title{Microfabrication of large area high-stress silicon nitride membranes for
optomechanical devices}

\author{E. Serra}
\affiliation{Istituto Nazionale di Fisica Nucleare, TIFPA, 38123 Povo (TN), Italy}
\affiliation{Delft University of Technology, Else Kooi Laboratory, 2628 Delft, The Netherlands}
\author{M. Bawaj}
\affiliation{\mbox{Physics Division, School of Science and Technology, Universit\`a di Camerino, 62032 Camerino (MC), Italy}}
\affiliation{INFN, Sezione di Perugia, 06123, Perugia, Italy}
\author{A. Borrielli}
\affiliation{Istituto Nazionale di Fisica Nucleare, TIFPA, 38123 Povo (TN), Italy}
\affiliation{\mbox{Institute of Materials for Electronics and Magnetism, Nanoscience-Trento-FBK Division, 38123 Povo (TN), Italy}}
\author{G. Di Giuseppe}
\affiliation{\mbox{Physics Division, School of Science and Technology, Universit\`a di Camerino, 62032 Camerino (MC), Italy}}
\affiliation{INFN, Sezione di Perugia, 06123, Perugia, Italy}
\author{S. Forte}
\affiliation{Delft University of Technology, Else Kooi Laboratory, 2628 Delft, The Netherlands}
\affiliation{Dipartimento di Fisica, Universit\`a di Trento, 38123 Povo (TN), Italy}
\author{N. Kralj}
\affiliation{\mbox{Physics Division, School of Science and Technology, Universit\`a di Camerino, 62032 Camerino (MC), Italy}}
\author{N. Malossi}
\affiliation{\mbox{Physics Division, School of Science and Technology, Universit\`a di Camerino, 62032 Camerino (MC), Italy}}
\affiliation{INFN, Sezione di Perugia, 06123, Perugia, Italy}
\author{L. Marconi}
\affiliation{\mbox{Dipartimento di Fisica e Astronomia, Universit\`a di Firenze, Via Sansone 1, 50019 Sesto Fiorentino (FI), Italy}}
\affiliation{INFN, Sezione di Firenze, Via Sansone 1, 50019 Sesto Fiorentino (FI), Italy}
\author{F. Marin}
\affiliation{LENS, Via Carrara 1, 50019 Sesto Fiorentino (FI), Italy}
\affiliation{\mbox{Dipartimento di Fisica e Astronomia, Universit\`a di Firenze, Via Sansone 1, 50019 Sesto Fiorentino (FI), Italy}}
\affiliation{INFN, Sezione di Firenze, Via Sansone 1, 50019 Sesto Fiorentino (FI), Italy}
\author{F. Marino}
\affiliation{INFN, Sezione di Firenze, Via Sansone 1, 50019 Sesto Fiorentino (FI), Italy}
\affiliation{CNR-INO, L.go Enrico Fermi 6, 50125 Firenze, Italy}
\author{B. Morana}
\affiliation{Delft University of Technology, Else Kooi Laboratory, 2628 Delft, The Netherlands}
\author{R. Natali}
\affiliation{\mbox{Physics Division, School of Science and Technology, Universit\`a di Camerino, 62032 Camerino (MC), Italy}}
\affiliation{INFN, Sezione di Perugia, 06123, Perugia, Italy}
\author{G. Pandraud}
\affiliation{Delft University of Technology, Else Kooi Laboratory, 2628 Delft, The Netherlands}
\author{A. Pontin}
\affiliation{\mbox{Dipartimento di Fisica e Astronomia, Universit\`a di Firenze, Via Sansone 1, 50019 Sesto Fiorentino (FI), Italy}}
\affiliation{INFN, Sezione di Firenze, Via Sansone 1, 50019 Sesto Fiorentino (FI), Italy}
\author{G.A.  Prodi}
\affiliation{Istituto Nazionale di Fisica Nucleare, TIFPA, 38123 Povo (TN), Italy}
\affiliation{Dipartimento di Fisica, Universit\`a di Trento, 38123 Povo (TN), Italy}
\author{M. Rossi}
\affiliation{\mbox{Physics Division, School of Science and Technology, Universit\`a di Camerino, 62032 Camerino (MC), Italy}}
\author{P.M. Sarro}
\affiliation{Delft University of Technology, Else Kooi Laboratory, 2628 Delft, The Netherlands}
%
\author{D. Vitali}
\affiliation{\mbox{Physics Division, School of Science and Technology, Universit\`a di Camerino, 62032 Camerino (MC), Italy}}
\affiliation{INFN, Sezione di Perugia, 06123, Perugia, Italy}
\author{M. Bonaldi}
\email[Corresponding author: ]{bonaldi@science.unitn.it}
\affiliation{Istituto Nazionale di Fisica Nucleare, TIFPA, 38123 Povo (TN), Italy}
\affiliation{\mbox{Institute of Materials for Electronics and Magnetism, Nanoscience-Trento-FBK Division, 38123 Povo (TN), Italy}}

\begin{abstract}

{In view of the integration of membrane resonators with more complex MEMS structures, we developed a general fabrication procedure for circular shape SiN$_x$ membranes using Deep Reactive Ion Etching (DRIE). Large area and high-stress SiN$_x$ membranes were fabricated and used as optomechanical resonators in a Michelson interferometer and in a Fabry-P$\acute{\textrm{e}}$rot cavity. The measurements show that the fabrication process preserves both the optical quality and the mechanical quality factor of the membrane.}
\end{abstract}


\maketitle

\begin{figure}[!t]
\includegraphics[width=86mm]{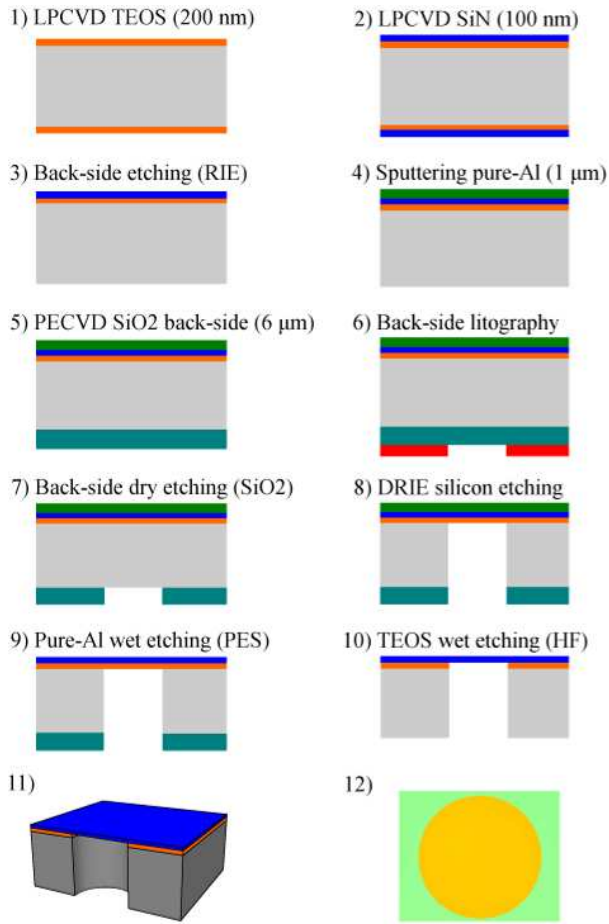}
\caption{1-10) Fabrication steps. 11) Section view of the finished device. 12) Optical microscope image of a circular membrane of diameter 1.5 mm.  
\label{fig:process}}    
\end{figure}

{The optomechanical coupling between a laser beam and a microdevice via radiation pressure \cite{optomechanicsRMP2014} is of great interest in quantum-optics and fundamental research. In fact, the latest micro- and nano-mechanical resonators, when used in a high-finesse Fabry-P$\acute{\textrm{e}}$rot optical cavity, offer great potential for precision sensing\cite{Purdy_Science_2013,bawajNComms2015} and for manipulation of the quantum state of light\cite{Safavi-Naeini_2013,Purdy_PRX2013}. 

In many cases the resonators consist of a free-standing high-stress silicon nitride (SiNx) membrane supported by a silicon (Si) frame, where Q-frequency product above $10^{13}$ Hz can be obtained\cite{WilsonPRL2009}  thanks to the large tensile stress (of the order of GPa).
These setups usually exploit dispersive coupling of the dielectric membrane placed in an optical cavity \cite{ThompsonNature2008}, but there are a number of ongoing efforts to extend the capabilities of SiN$_x$ membrane resonators, for instance by coating them with a metal for use in hybrid optical-microwave setups \cite{Andrews2014} or by enhancing optomechanical coupling by patterning of photonic crystal structures\cite{bui}. We also mention recent studies aiming to understand and possibly overcome the current limits in their mechanical performance \cite{VillanuevaPRL2014}.

Large area  free-standing silicon nitride membranes were originally proposed as TEM windows.
Generally, these are fabricated on a silicon (Si) support by low pressure chemical vapour deposition (LPCVD) and then released by wet etching the Si substrate. For this last step  potassium hydroxide (KOH) solutions are typically employed. However, this etching is highly selective along silicon crystal planes and allows precise control of dimensions only if the desired structure can be bounded by $<111>$ planes \cite{Leondes}, as in rectangular membranes. On the other hand the integration of the resonator in complex microsystems requires a greater flexibility in terms of layout. 

In this paper we describe a general fabrication procedure based on Deep Reactive Ion Etching (DRIE) through-wafer etching. The use of DRIE is really advantageous as it allows the fabrication of any-shape membranes. For example circular membranes have been shown to have superior mechanical properties\cite{Wilson-RaePRL2011}. At the same time DRIE enables the fabrication of complex hybrid system around the membrane. This is required to ease the integration of the SiN$_x$ membrane with on-chip mechanical isolation, useful for improving the overall reproducibility of the resonator by reducing the mechanical coupling with the sample holder. We remark that this strategy has recently allowed the production of optomechanical microresonators where quality factors higher than 10$^6$ can reproducibly be obtained for specifically designed normal modes \cite{SerraAPL2012,BorrielliPRApplied2015}. So far SiN$_x$ resonators suspended from phononic band-gap isolation structures have been built by combining DRIE with a wet etch release of the SiN$_x$ layer, but the process could not attain an optimal control of the membrane cleanliness \cite{YuAPL2014,TsaturyanOE2014}.
In our case, a complete characterization shows that the main mechanical and optical features of the resonator are preserved for a lot of circular membranes. 

Membranes were fabricated following the steps shown in Figure \ref{fig:process}. As substrate we employed a double-side-polished Si wafer with a thickness of 500 $\mu$m and a RMS surface roughness lower than 1 nm. 
After standard cleaning by $\rm 99 \%$ and $\rm 65 \%$  $\rm H NO_3$ baths followed by DI-water rinse steps, a LPCVD 200-nm-thick  tetraethylorthosilicate (TEOS)  SiO$_2$ layer was deposited on the silicon substrate (step 1). This layer has a low compressive stress and is used as etching stop for the DRIE process.
A 100-nm-thick LPCVD nitride ($\rm SiN_x$) layer was then deposited (step 2) after optimizing\cite{Sarro} the recipe for a residual tensile stress of about 1 GPa having a refractive index at 1064 nm of $\rm n=2.021$.
These properties were obtained by laser-based wafer curvature measurements (Flexus FLX-2908) and by variable angle spectroscopic ellipsometry (VASE), respectively. 
After the back-side etching (step 3),  a low-stress 1-$\rm \mu m$-thick layer of pure $\rm Al$ was sputtered at $25 \;^{\circ}{\rm C}$ (step 4) to protect the $\rm SiN_x$  during the DRIE step. 
We used a plasma enhanced chemical vapour deposition (PECVD) 6-$\rm \mu m$-thick $\rm SiO_2$ layer (step 5), patterned with circular holes (step 6-7), as masking layer for the substrate etching. 
The Si was then removed (step 8) by means of DRIE (Omega i2L Rapier).
The resulting etching rates were about 1.38 $\rm \mu m/cycle$ for $\rm Si$ and 5 $\rm nm/cycle$ for the oxide layer. To release the SiN$_x$ layer we first stripped the $\rm Al$ layer (step 9) by a PES   solution, followed by a dip etching (step 10) in HF solution. The wafer is then rinsed in DI-water and diced.
}

\begin{figure}[h]
\centering
\includegraphics[width=86mm]{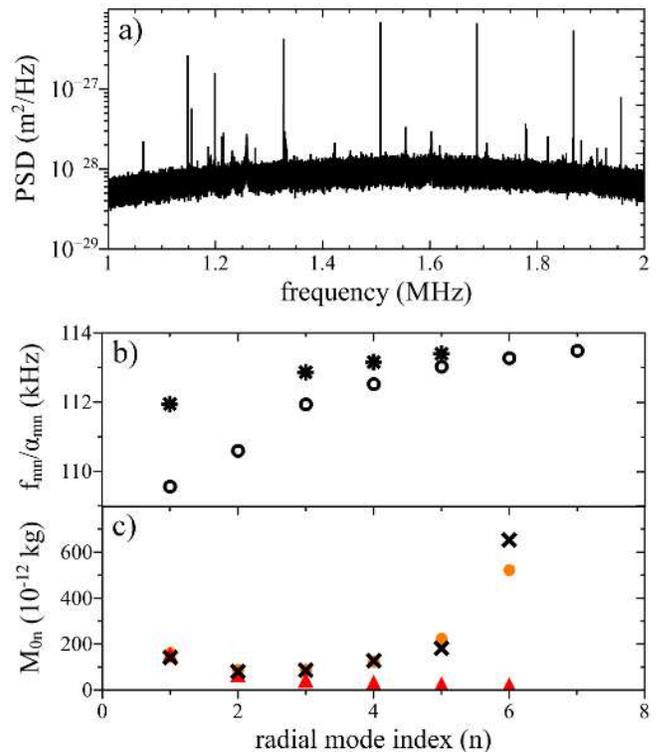}
\caption{a) Typical noise power spectral density (PSD) of the interferometer output; the modal frequencies can be clearly seen above background noise. b) Resonance frequencies of the first $0n$ (open circles) and $1m$ (stars) modes af a $R=0.75$ mm circular membrane, divided by the respective $\alpha_{mn}$. c) Experimental effective mass $M_{0n}$ of the first modes (crosses), theoretical values calculated for a centered Gaussian readout with width $w=0.15$ mm (closed circles) and for a pointlike readout (triangles).
\label{fig_f0}}
\end{figure}

We focus the analysis of the mechanical resonance frequencies on the circular membranes, due to their original shape with respect to the more common square membranes. The theoretical resonance frequencies in a circular membrane are given by the expression
$
f_{mn} = f_0\,\alpha_{mn}
$
where $\alpha_{mn}$ is the $n$-th root of the Bessel polynomial of order $m$, and
$
f_{0} = \frac{1}{2 \pi}\, \sqrt{\frac{T}{\rho}}\frac{1}{R}
$
($T$ is the stress, $\rho$ the density, $R$ the radius of the membrane). We have measured the modal frequencies from a thermal spectrum acquired using a Michelson interferometer (Fig. \ref{fig_f0}a), with the sample kept in a vacuum chamber. In Fig. \ref{fig_f0}b we show the experimental resonance frequencies divided by the respective $\alpha_{mn}$, for the first $m=0$ and $m=1$ order modes. We expect a constant value, equal to $f_0$. It appears that the lower modes slightly (but systematically) deviate from the predicted behavior, with a maximum spread of just $4\%$, probably due to boundary effects, likely influenced by the clamping. We can extrapolate an asymptotic experimental value of $f_0 \simeq 114$ kHz, to be compared with $f_0 = 118.6$~kHz that is calculated using the nominal parameters $R = 0.75$~mm, $T = 1$~GPa and $\rho = 3200$~kg/m$^3$. The agreement is very good. We also remark that measurements taken at different times (entailing few degrees variations of the room temperature) yield fluctuations of the experimental $f_0$ by $2\div3 \%$.

An interesting property of the circular membrane is that the effective modal mass depends on the modal index, as opposed to the square membranes. Namely, the effective mass of the $0n$ modes for a centered,  $\delta-$like readout is $M_{0n} = M \left(J_1\left(\alpha_{0n}\right)\right)^2$, where $M$ is the physical mass of the membrane and $J_m$ is the Bessel polynomial (the first values of $M_{0n}/M$ are 0.269, 0.116, 0.074, 0.054, 0.043). The mass is lower at higher index because the modal displacement is more concentrated in the center, while it remains homogeneous for a square membrane. For comparison, in the case of a square membrane the effective mass of the odd modes is $M/4$. A reduced effective mass, increasing the susceptibility, is a useful property in opto-mechanics experiments. In the realistic case of a centered, Gaussian readout with 1/e$^2$ width $w$, the effective mass becomes:
\begin{equation}
M_{0,n} = M \left(\frac{J_1\left(\alpha_{0n}\right)}{\frac{4}{w^2}\int J_0 \left(\alpha_{0n}r/R\right) \exp \left(\frac{2 r^2}{w^2}\right) r \mathrm{d}r}\right)^2  \, .
\label{Meff}
\end{equation}
In Fig. \ref{fig_f0}c we report the experimental values of $M_{0,n}$ for the first modes, derived from the areas $A_{0n}$ of the thermal peaks in the displacement spectrum using $A_{0n}= \frac{k_B \mathrm{T}_K}{M_{0m} (2 \pi f_{0n})^2}$ ($k_B$ is the Boltzmann constant and T$_K$ the temperature). They are compared with the theoretical values calculated for a realistic $w=0.15$~mm (showing a good agreement) and with a pointlike readout.

The mechanical quality factor $Q$ of the modes of different membranes has been measured both at room and at cryogenic temperatures, by driving the different resonances with a piezoelectric glued on the sample mount, and observing the ring-down with a Michelson interferometer. At room temperature, the $Q$ values are very scattered, ranging from few thousands up to $2\times 10^5$. This feature is very common in SiN$_x$ membranes, and is due to the coupling with the frame and, through it, with the sample holder. At cryogenic temperatures, the values of $Q$ are still scattered, but globally higher. With a 1.5~mm diameter membrane on a 5~mm side, square frame clamped between two copper plates, we could measure a maximum $Q$ of $0.65\times10^6$, at 8~K. With a 1~mm side, square membrane close to the edge of a $5 \times 20$~mm$^2$ frame, glued on the opposite edge to a copper block, we could measure $Q$ values up to $1.3 \times 10^6$ at 13~K. Such results are comparable with those commonly obtained with commercial, high-stress SiN$_x$ membranes and confirm the validity of our fabrication procedure, while stressing the importance of engineered frames and holders.

\begin{figure}[!t]
\includegraphics[width=86mm]{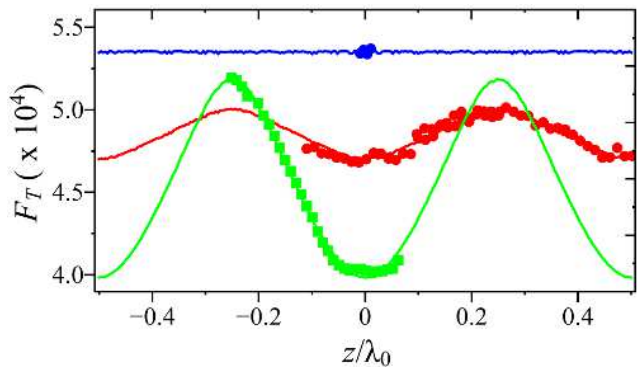}
\caption{Plot of the cavity finesse, ${\mathcal F}_T$ versus membrane position along the cavity axis.
    Red-symbols represent data for a circular-membrane, 1.2~mm diameter, and 97~nm thickness.
    Red-line represents the best fit with fitting parameters $n_I = (1.97\pm 0.08)\times 10^{-6}$ and {$\sigma_{\textrm{opt}} =(287 \pm 4)\,$}pm, and with the following fixed parameters: the membrane thickness $L_d = 97$~nm, the real part of the {refractive index} $n_R = 2.021$, the cavity length $L = 9.03$~cm, the wavelength $\lambda = 1064$~nm, and the empty-cavity finesse ${\mathcal F}_v = 53518$, which is evaluated as mean value of the data shown as blue-symbols.
    Green-symbols represents data for a square-membrane, 1~mm side, 50~nm thickness by Norcada, The green-line is the fitting with parameters {$n_I = (1.0 \pm 0.01)\times 10^{-5}$} and {$\sigma_{\textrm{opt}} =  (280 \pm 10)\,$}pm.
\label{fig:Finesse_Scattering_FIT}   }    
\end{figure}

We characterized the optical properties of a circular membrane of nominal radius $R = 0.6$~mm by first performing VASE measurements soon after its fabrication, obtaining the following values for the membrane thickness, $L_d = 97.27 \pm 0.01$ nm, index of refraction at $\lambda = 1064 $ nm $n_R = 2.0210 \pm 0.0005$. 
These values have been confirmed by polarization-resolved transmission measurements again at $\lambda = 1064 $ nm, performed on the final, diced samples. These latter measurements provide an intensity reflectivity at normal incidence $ |r_d|^2 = 0.355 \pm 0.002$, perfectly consistent with the expression for the amplitude reflection and transmission coefficients of a dielectric membrane of thickness $L_d$ and refractive index $n$,
\begin{eqnarray} \label{eq:r_d}
	r_d =  \frac{(n^2 -1)\sin\beta}{2in\cos\beta + (n^2 +1)\sin\beta},  \\
	t_d = \,\frac{2n}{2in\cos\beta + (n^2 +1)\sin\beta}\,,\label{eq:t_d}
\end{eqnarray}
where $\beta = n kL_d = 2\pi n L_d/\lambda$. When $n=n_R$ is real there is no optical absorption, but in general $n = n_R+i n_I$, with the small but nonzero imaginary part $n_I$ providing a measure of optical absorption. In order to accurately evaluate it, we have placed
the membrane between two Al cylinders, and we have inserted it in a $L=9.03$~cm long cavity at room temperature formed by two spherical mirrors of radius of curvature $7.5$~cm.
The high-frequency component of the Pound-Drever-Hall (PDH) signal used to lock the cavity was acquired for the preliminary analysis of the mechanical modes at a vacuum chamber pressure of $1.7\times10^{-2}$~mbar. The fundamental eigenfrequency is found to be $\nu_{01}\sim 348.5$~kHz, consistent with the theoretical expectation if we assume $R = 0.615$~mm.
The detected optical cavity modes as a function of the membrane position are also consistent with the theoretical model~\cite{Jayich:2008nx,Biancofiore:2011uq} obtained with the cavity and membrane parameters. 
Finally the cavity finesse, ${\mathcal F}_T$, as a function of the membrane position has been measured (see Fig.~\ref{fig:Finesse_Scattering_FIT}) by the ringdown technique. The laser beam, deflected by an acousto-optic-modulator (AOM) and sent into the cavity, is switched off in 50~ns after the transmitted light has reached a threshold level. The leakage of transmitted light is monitored to estimate the cavity decay-time, which is fitted with a single exponential form whose time constant, $\tau$, is related to ${\mathcal F}_T$ via ${\mathcal F}_T = \pi c \tau/ L$. Repeating the finesse measurement for different positions of the membrane as reported in Fig.~\ref{fig:Finesse_Scattering_FIT}, allows to estimate $n_I$ and the roughness of the membrane.
For fitting the finesse data we consider the transfer function of a cavity with a membrane in the middle. The cavity consists of two semi-cavities;
denoting with $z$ the position of the membrane with respect to the cavity center, and assuming for the reflection and transmission coefficients of the mirrors $r_1 = r_2 =\sqrt{{\mathcal R}}$ and  $t_1 = t_2 = \sqrt{{\mathcal T}}$, we derive the following expression for the intensity transmission of the cavity,
\begin{equation}\label{transm}
{\mathcal T}_c  = \frac{\left|{\mathcal T}t_d\right|^2}{\left|1+2r_d\sqrt{{\mathcal R}} \cos( 2kz){\rm e}^{ikL} + {\mathcal R}(t_d^2 +r_d^2) {\rm e}^{2ikL}\right|^2}\,.
\end{equation}
In the ideal case ${\mathcal R} = 1$ and $n$ is real, so that $t_d^2 +r_d^2 = \exp\left\{2i\arg(r_d)\right\}=\exp\left\{2i\phi_r\right\}$, the eigenfrequencies of the cavity modes are given by the zeros of the denominator of the cavity transmission
\begin{equation}\label{eq1}
[\cos(kL +\phi_r) + |r_d| \cos( 2kz)]^2 =  0\,,
\end{equation}
which gives the results in Ref.~\cite{Jayich:2008nx}
\begin{equation}\label{eq2}
2kL  = \nu\frac{2\pi}{\Delta\nu_{FSR}}= - 2\phi_r + 2\cos^{-1} [-|r_d| \cos( 2kz)]\,.
\end{equation}
When $\mathcal{R} < 1$ and in the presence of absorption the resonance frequencies of the cavity modes are determined by the maxima of the transmission ${\mathcal T}_c $ given by Eq.~(\ref{transm}), and the finesse can be obtained from the width of the transmission peaks.
Roughness can be introduced following Refs.~\cite{Kleckner:2010,Winkler:1994,virgo}, according to which any optical element with a given roughness is responsible for a modification of the wavefront and therefore leads to a scattering loss from the incident optical mode into all the other cavity modes. This scattering into other modes induced by an {effective optical roughness $\sigma_{\textrm{opt}}$ can be modeled by including a factor $\sqrt{\exp[-(2k\sigma_{\textrm{opt}})^2]}$ multiplying $r_d$ in eq.~\eqref{eq:r_d}.}
Fig.~\ref{fig:Finesse_Scattering_FIT} shows the best fits for the set of data (red symbols) taken for the 1.2~mm--diameter circular membranes. We have taken $n_I$ and {$\sigma_{\textrm{opt}}$} as fitting parameters, while we have kept fixed all the other parameters ($L_d = 97$~nm, $n_R = 2.021$, $L = 9.03$~cm, $\lambda = 1064$~nm, and ${\mathcal F}_v = 53518$ corresponding to ${\mathcal R}=0.9999413$).  Both absorption and roughness give a $z$-dependent effect, with a $\lambda/4$ periodicity. The fit gives an imaginary part of the {refractive index} $n_i = (1.97\pm 0.08)\times 10^{-6}$ and {$\sigma_{\textrm{opt}} =(287 \pm 4)\,$}pm. As a comparison, the results for a 1~mm side, 50~nm thickness membrane by Norcada~\cite{norcada} are reported with fitting parameters {$n_I = (1.0 \pm 0.01)\times 10^{-5}$} and {$\sigma_{\textrm{opt}} =$  $(280 \pm 10)\,$}pm. The circular membranes we fabricated present {a similar effective optical roughness with respect to a commercial square one but a lower absorption. As regards standard AFM measurement of roughness, we measured for our membranes $\sigma_{\textrm{rms}}= 0.7 \pm 0.1\,$nm over scan areas of 1$\mu$m$\times 1\mu$m.}

In conclusion, we developed and validated a general fabrication procedure for free-standing large area high-stress SiN$_x$ membranes of any shape with a good dimensional precision by using DRIE etching. 
Possible improvements of the process could be obtained by the use of silicon substrates with a lower initial roughness. This work is a crucial step toward the integration of SiN$_x$ membranes in hybrid systems and on-chip mechanical isolation.

\end{document}